\documentclass[conference, a4paper, 10pt]{IEEEtran}
\IEEEoverridecommandlockouts
\usepackage{booktabs}
\usepackage{multirow}
\usepackage[super]{nth}

\usepackage[moderate, tracking=normal]{savetrees}

\usepackage{amsmath,amssymb,amsfonts}
\usepackage{algorithmic}
\usepackage{graphicx}
\usepackage{textcomp}
\usepackage{xcolor}
\usepackage[parse-numbers=false,binary-units=true]{siunitx}
\def\BibTeX{{\rm B\kern-.05em{\sc i\kern-.025em b}\kern-.08em
    T\kern-.1667em\lower.7ex\hbox{E}\kern-.125emX}}
    
\graphicspath{{fig/}}
\usepackage{url}

\usepackage{caption}
\usepackage{varioref}

\usepackage{array}
\newcolumntype{P}[1]{>{\raggedright\arraybackslash}p{#1}}
\newcolumntype{C}[1]{>{\centering\arraybackslash}p{#1}}

\usepackage{verbatim}
\usepackage[nottoc]{tocbibind}
\usepackage[numbers]{natbib}

\newcommand{\B}[1]{\boldsymbol{#1}}

\usepackage{todonotes}

\begin{document}

\title{Neural Network Quantisation \\ for Faster Homomorphic Encryption
\thanks{This project has received funding from the European Research Council (ERC) under the European Union's Horizon 2020 research and innovation programme (grant agreement nr. 101020005) and from the
Defense Advanced Research Projects Agency (DARPA) under Contract No. HR0011-21-C-0034. Michiel Van Beirendonck is funded by FWO as Strategic Basic (SB) PhD fellow (project number 1SD5621N). Jan-Pieter D’Anvers is funded by FWO (Research Foundation – Flanders) as junior post-doctoral fellow (contract number 133185).
}
}


\author{
    \IEEEauthorblockN{
     Wouter Legiest, Furkan Turan, Michiel Van Beirendonck, Jan-Pieter D'Anvers and Ingrid Verbauwhede}
    \IEEEauthorblockA{
    \textit{COSIC, KU Leuven}, Belgium \\
    \texttt{\{firstname.lastname\}@esat.kuleuven.be} \\} 
    \vspace{-1em}
}

\IEEEoverridecommandlockouts \IEEEpubid{\makebox[\columnwidth]{979-8-3503-4135-5/23/\$31.00~\copyright2023 IEEE \hfill} \hspace{\columnsep}\makebox[\columnwidth]{ }}

\maketitle

 \IEEEpubidadjcol

 


\begin{abstract}
Homomorphic encryption (HE) enables calculating on encrypted data, which makes it possible to perform privacy-preserving neural network inference. One disadvantage of this technique is that it is several orders of magnitudes slower than calculation on unencrypted data. Neural networks are commonly trained using floating-point, while most homomorphic encryption libraries calculate on integers, thus requiring a quantisation of the neural network. A straightforward approach would be to quantise to large integer sizes (e.g. \SI{32}{\bit}) to avoid large quantisation errors. 
In this work, we reduce the integer sizes of the networks, using quantisation-aware training, to allow more efficient computations. For the targeted MNIST architecture proposed by \citet{BJLMJTNAC21}, we reduce the integer sizes by 33\% without significant loss of accuracy, while for the CIFAR architecture, we can reduce the integer sizes by 43\%. Implementing the resulting networks under the BFV homomorphic encryption scheme using SEAL, we could reduce the execution time of an MNIST neural network by 80\% and by 40\% for a CIFAR neural network.

\end{abstract}


\section{Introduction}

Homomorphic encryption (HE) allows performing calculations on encrypted data. This technique enables applications where data is processed in untrusted environments (e.g. a cloud environment) while ensuring that this environment does not learn anything about the data itself. As such, it is a promising technique to make privacy-preserving machine learning possible. 

A downside of HE is that it significantly increases the size of encrypted data. As a result, encrypted operations are typically several orders of magnitude slower than their unencrypted counterparts. This work tries to accelerate neural network inference under homomorphic encryption by using quantisation techniques to reduce the data size and, thus, the computational cost. 

Neural network frameworks generally use a floating-point representation to represent network parameters and intermediate variables. However, HE systems like BFV~\cite{FV12} encode only integers, requiring an additional conversion step to convert the floating-point neural network parameters to the integer HE variables. While it is possible to design neural networks that work solely with integer representations, previous works have only studied such networks in a non-HE related context~\cite{jacob2017quantization, Moons2016, electronics10222823}. 

In addition, this conversion is an essential step before porting it to hardware. For instance, a plaintext \SI{32}{\bit} floating-point addition is $30\times$ more energy-consuming \footnote{Energy consumption using a \SI{45}{\nano\meter} CMOS technology.} than an \SI{8}{\bit} integer equivalent~\cite{horowitz}.
Using the conversion, we can select smaller HE parameters that lead to limited resource use and better management of corner cases. Therefore, making the behaviour of the system faster and more predictable in general. 

As calculations in these non-HE integer-only networks are performed, the sizes of the integer variables increase. The intermediate values are commonly scaled down to a smaller number after each layer to keep these integer-only networks manageable. This means the most significant bits are held after each operation, while the least significant are discarded. 

Unfortunately, these reduction operations are based on division or shift operations, which natively are not supported in HE schemes, so downscaling cannot easily be performed.
Therefore, in neural network HE inferences, the intermediate values will grow throughout the inference and the final calculations will need to operate on very large integers.
For instance, when all of the weights of a neural network are converted to \SI{32}{\bit}, a 10-layer CIFAR network will produce integers with bit-sizes up to \SI{614}{\bit}.
The maximum bit-length of these output integers will be denoted as the \emph{`final integer width'} (FIW), and we will show that this value significantly affects the overall computation cost.

\citet{GD16} implemented the first artificial feedforward neural network under homomorphic encryption using the HE scheme YASHE~\cite{Yashe}.
Note that an attack proposed by~\citet{YASHE_attack} reduced the security level of this scheme and is therefore considered broken in practice.
\citet{GD16} proposed a specialised, HE-focussed \emph{CryptoNets} architecture for the MNIST dataset~\cite{LBBH98}. 

One of the downsides of the CPU implementations of CryptoNets is the high latency of \SI{250}{\sec} for an MNIST image.
It was improved by \citet{BGE19} with the Low-Latency CryptoNets (LoLa) architecture. Using the BFV scheme, optimisations in the underlying HE library SEAL and a different approach to representing the ciphertext data, a latency of \SI{0.29}{\sec} was reached, an improvement of $93\times$ relative to CryptoNets. 
In addition, \citet{BGE19} proposes variants of the LoLa network for processing the CIFAR-10 dataset~\cite{Krizhevsky_2009_17719}. They report an accuracy of 74.1\% and a latency of \SI{730}{\sec}.

\citet{BJLMJTNAC21} implemented the BFV scheme on GPUs. They propose two architectures, one smaller for MNIST and one more extensive for CIFAR. Accordingly, their CIFAR network boasts an accuracy of 77.55\% and a latency of \SI{304.43}{\sec}. 

In this work, we improve upon the state-of-the-art HE neural networks by considering advanced neural network quantisation techniques. We first investigate post-training quantisation, a method typically used in the state-of-the-art, and show that there is a limit to how many intermediate variables can be scaled down without significantly affecting accuracy. We then show that quantisation-aware training can indeed be used to substantially scale down these intermediate variables without a similar accuracy penalty. In the end, we reduced the final integer width with 33\% for MNIST and 43\% for CIFAR, allowing a speedup with factors 80\% and 40\%, respectively, over typical 8-bit post-training quantisation networks as used in the state-of-the-art.


\section{Preliminaries}


\subsection{Homomorphic encryption}

Homomorphic encryption enables the performing of arithmetic operations on encrypted data. 
Take the following example: consider an asymmetric encryption system with two integer numbers $x$ and $y$. They can be encrypted by the using encryption key $\mathsf{pk}$ to $c_x=\mathsf{Enc}(\mathsf{pk},x)$ and $c_y=\mathsf{Enc}(\mathsf{pk},y)$. These two ciphertexts are sent to a server that cannot be trusted. The server can perform an operation $\diamond$ on both ciphertext $c_{xy} = c_x\diamond c_y$, which is the equivalent of doing an addition on the plaintexts. The result of this operation is then sent back to the user, who can decrypt this message. Using the decryption key $\mathsf{sk}$, the resulting plaintext message $z$ is obtained by $z = \mathsf{Dec}(\mathsf{sk},c_{xy})$. The message $z$ will have a value of $z = x + y$. Altogether, the server does not obtain any information about the integers $x$ and $y$ while possessing the unencrypted data. 

A limitation of this form of encryption is that it only allows certain operations, i.e. addition or multiplication of two ciphertexts. Execution of non-linear functions is normally performed using a polynomial approximation that uses only addition, subtraction, and multiplication. Moreover, a division in the HE schemes CKKS and BFV is theoretically possible, but it is costly and thus avoided in practice~\cite{DNW17}.

The biggest problem with the lack of a division operation is that variables will grow during computation. For example, when multiplying two \SI{8}{\bit} integers, the result becomes roughly \SI{16}{\bit}. In unencrypted neural network implementations, this variable can be divided with a power of two to get back to \SI{8}{\bit}, making it more manageable for the next layer. However, such an operation is not possible in encrypted neural network inference. This leads to large intermediate and output integers. The maximum bit-length of these output integers will be denoted as the `final integer width' (FIW).

The HE scheme must be instantiated with more extensive parameters to accommodate these larger variables, which comes at a significant cost. 
Once a specific variable size is reached, additional techniques are required to allow large representations.
More specifically, to ensure a correct representation in the plaintext space during inference, a residue numeral system (RNS), following the Chinese remainder theorem, is used to divide the large numbers into several smaller numbers. 
This leads to several smaller HE instances that could be run in parallel. 
Since each instance consumes computing resources, decreasing the variable sizes can significantly reduce the number of RNS instances and, thus, the computational cost.

\subsection{Neural network}

A neural network is a machine learning technique consisting of a network of small interconnected computation units called neurons. These neurons can be adapted, which enables the network to `learn' a specific, human-like task such as classifications of images. A neuron will take a number of inputs, perform a weighted sum over these inputs, and output a function of the result of this sum.
Neurons are grouped to form layers, and different behaviour can be obtained depending on their configuration.

Since a typical division or non-linear function cannot be executed trivially under FHE, we use a slightly adapted version of the classical neural network layers. A dense or convolutional layer is representable under FHE. However, the activation function is approximated by a $f(x) = x^2$ square function, resulting in a \emph{Square layer}. Moreover, the \emph{scaled average pooling} layer is replaced by an equivalent where the inputs are summed, but the division is omitted.



\subsection{Architectures}

This work uses the two architectures developed by \citet{BJLMJTNAC21} for homomorphic inference. These networks are used as test cases to research the effect of quantisation on homomorphic encryption inference. Both architectures omit the last (Sigmoid) activation function since it only maps the output to the unit interval.  
For a detailed description of the network, we refer the reader to the paper of \citet{BJLMJTNAC21}.  

The first architecture used in this paper focuses on the MNIST dataset~\cite{LBBH98}. 
It is based on the HCNN~\cite{BJLMJTNAC21} and consists of two convolutional, two square activation and one dense layer. 
The authors stated an accuracy of 99\% for this architecture. 

The second architecture is designed to classify the more complex CIFAR-10 dataset~\cite{Krizhevsky_2009_17719}. The 10-layered network especially uses the scaled average pooling and square layer.
The proposed initial HCNN architecture was slightly modified in our implementation by not using padding, as this only results in reduced accuracy: our floating-point model 
obtains an accuracy of 73.28\%, while the original HCNN reports 77.8\%. 

\section{Post-Training Quantisation (PTQ)}

Usually, floating-point numbers with single or double precision are used to represent the weights and biases of a network. However, it is possible to convert these numbers to \SI{8}{\bit} integers without a notable reduction in accuracy~\cite{han2016deep}. A further reduction in the representation might have a more detrimental effect on the neural network accuracy. 
Converting an existing (floating-point) neural network into a quantised (integer) version is called post-training quantisation (PTQ). 

During PTQ quantisation, a real value $r \in [\alpha, \beta]$ is converted to a $b$-bit integer $q$. The process is determined by two factors: the zero-point $Z$ and the scale factor $S$, using the following formula:
\vspace{1.5em}
\begin{align} 
\label{eq_quan} 
q &= \left\lfloor \frac{r}{S} + Z\right\rceil. 
\end{align}
\vspace{0.5em}

Dequantisation can be done through the formula $r = S(q - Z)$, where the quantised value is converted back to its original scale.


The scale factor $S$ determines the quantisation step size. 
The zero-point $Z$ is the quantised value $q$ corresponding to the real value $r=0$ and positions the range of representable numbers optimally.

When $Z \neq 0$, we say the quantisation is asymmetric or affine. This quantisation explicitly uses the zero point, often set at $Z=-\alpha\cdot(2^b-1)/(\beta-\alpha)$. A second option is a symmetric quantisation, which reduces the overhead of dealing with the zero point by setting it to zero. 
Commonly, the values are mapped to a signed symmetric interval $[\alpha,\beta] = [-2^{b-1}, 2^{b-1}-1]$, although an unsigned interval is also possible. Symmetric quantisation is a more limited but easier-to-handle quantisation technique.  

\begin{figure}[!htb]
  \centering
      \includegraphics[width=0.5125\columnwidth]{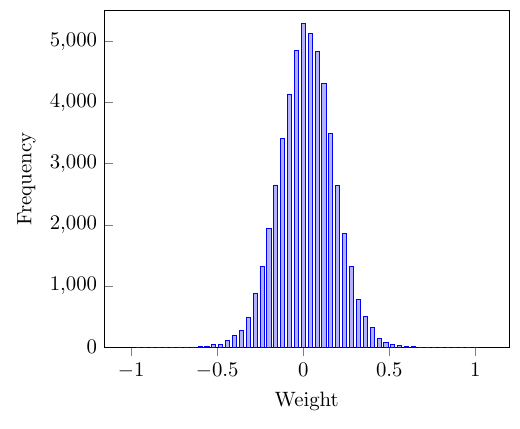}
      \includegraphics[width=0.478\columnwidth]{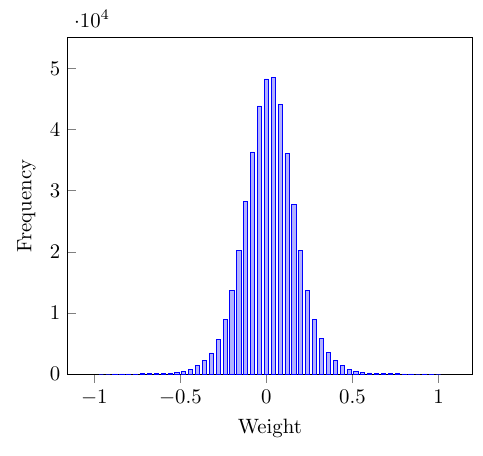}
  \caption{Overview of the weight distribution of the MNIST and CIFAR architecture.}
  \label{fig_hist}
\end{figure}

To evaluate the effect of quantisation, we first determined the distribution of the parameters of both MNIST and CIFAR networks, plotted in Figure~\ref{fig_hist}. Since both networks possess a symmetric distribution, the mean of the values is zero, and thus symmetric quantisation is the best candidate to convert the signed real numbers for both networks. To determine the ideal scale factor for the weights, three candidates are tested.

The first scale factor $S = 1/(2^{b-1}-1)$ only considers the bit width. No account is taken of the size or distribution of the real numbers. The second scaling factor $S = \max(|\B{W}|)/(2^{b-1}-1)$ considers the largest absolute value that can be represented in the quantised interval, and extreme values that are outside the quantisation range are quantised to the edges of the quantisation interval. 
The third factor $S = (\beta - \alpha)/(2^{b-1}-1)$, uses the length of the interval.  This way, the length of the real interval is divided by the length of the quantised interval. This form is typically used for asymmetric quantisation.

To understand the influence of these different scale factors on the network, we build a Python framework that evaluates the effect of post-training quantisation on the accuracy and FIW. The framework takes a neural network, converts each of the weights to an integer representation and then executes a neural network inference. We process each of the $10\;000$ images in the test set for these experiments to determine the accuracy and FIW. The maximum of all individual final integer widths and corresponding accuracies are reported in Table~\ref{tab_ptq}. 



Furthermore, we also reduce the sizes of the input coefficients. This way, we can obtain an even lower final integer width. In all the experiments, the MNIST data is scaled down from its typical \SI{8}{\bit} to \SI{2}{\bit}. However, since CIFAR images are more complex, the same reduction could lead to unacceptable accuracies. Therefore we chose not to reduce the CIFAR dataset. 

The results in Table~\ref{tab_ptq} show that for both networks, we can quantise until \SI{8}{\bit} without an accuracy drop. When we want to use lower quantisation, the accuracy starts to drop.
One of the reasons is that in these cases, many of the weights are quantised to zero, which causes much of the information to `disappear' and results in a diminished FIW. 

\begin{table*}[!ht]
\centering
\caption{Results of the quantised model using Brevitas and post-training quantisation with different scale factors for the architectures for both MNIST and CIFAR.}
\label{tab_ptq}
\begin{tabular}{@{}llcccccccc@{}}
\toprule
        &             & \multicolumn{2}{c}{PTQ -- $S =\frac{\max(|\B{W}|)}{2^{b-1}-1}$} & \multicolumn{2}{c}{PTQ -- $S =\frac{\beta-\alpha}{2^{b-1}-1}$} & \multicolumn{2}{c}{PTQ -- $S =\frac{1}{2^{b-1}-1}$}         & \multicolumn{2}{c}{QAT} \\ \cmidrule(l){3-4} \cmidrule(l){5-6} \cmidrule(l){7-8} \cmidrule(l){9-10}
Dataset & Quantisation & Acc [\%]                        & FIW [bit]          & Acc [\%]                & \multicolumn{1}{c}{FIW [bit]}  & Acc [\%]              & \multicolumn{1}{c}{FIW [bit]} & Acc [\%]   & FIW [bit]  \\ \midrule
   & \SI{32}{\bit}          & 98.43                           & 237                & 98.43                        & 231                               &             98.43          & 233                              & 98.6      & 238        \\
\multirow{2}{*}{MNIST}   & \SI{8}{\bit}          & 98.41                    & 69                & 98.44                  & 63   & 98.44                      & 64                              & 98.51      & 70         \\
        & \SI{3}{\bit}             & 94.39                           & 32                 & 44.03                      & 26                               & 79.04                      & 27                              & 98.3       & 38         \\
        & \SI{2}{\bit}             & 14.4                            & 20                 & 11.35                        & 3                               & 11.53                      & 6                              & 98.46      & 29         \\ \midrule 
   & \SI{32}{\bit}            & 73.09                           & 583                & 73.09                   & 571                            & 73.09                      & 570                              & 73.28      & 614        \\
CIFAR        & \SI{8}{\bit}             & 73.0                            & 202                & 73.18                   & 187                            & 73.09                     & 186                             & 73.04      & 205        \\
        & \SI{4}{\bit}             & 52.24                           & 135                &  18.67                       & 123                               & 9.96                      & 128                             & 72.49      & 153        \\
        \bottomrule
\end{tabular}
\end{table*}

\section{Quantisation-aware training (QAT)}

In the previous section, we showed that neural networks could be quantised to \SI{8}{\bit} integers, but the accuracy reduces for a more drastic quantisation. To reduce the bit width of the network further, we can make the network aware of the quantisation during its training. 
Before training is started, the quantisation technique and parameters are chosen and introduced into the training graphs as `fake quantisation' nodes, which simulate the low-precision behaviour of the quantisation. 
These nodes quantise a real input using Equation~\ref{eq_quan} and perform a dequantisation immediately afterwards, thus injecting an error that the quantisation would cause. Depending on the quantisation used, this method can result in a network with approximately the same accuracy as a full-precision network while using low-precision parameters. 

Using Brevitas~\cite{brevitas}, we trained the same networks as in the post-training quantisation experiments of the previous section. Brevitas is a library to develop and train quantisation-aware hardware-ready networks. We used its `weights-only quantisation process', in which a quantisation error is exclusively injected in the weights.

The results can be seen on the right in Table~\ref{tab_ptq}. 
For lower bit widths, the accuracy of the QAT is significantly better than the PTQ case, remaining approximately the same as the full precision network. One of the reasons is that quantisation-aware training will prevent parameter sparsity and ensure that each parameter is used correctly.

\begin{table}[!ht]
\caption{Results of the quantisation-aware training and homomorphic inference.}
\label{tab_QAT}
\centering
\begin{tabular}{P{0.1cm}P{1.1cm}P{0.1cm}P{1cm}P{0.75cm}P{0.5cm}|P{0.75cm}P{0.75cm}}
\toprule
                       & Network && Quant              & Acc. [\%] & FIW [\SI{}{\bit}] & Seq. time [min.] & No. of inst. \\ \midrule
      & CryptoNets &\cite{GD16}  & 5-\SI{10}{\bit} & 99    & 80  & \multicolumn{1}{c}{-}     &  2  \\
\multirow{5}{*}{\rotatebox{90}{MNIST}} & HCNN & \cite{BJLMJTNAC21}                 & \SI{4}{\bit}               & 99    & 43  & \multicolumn{1}{c}{-}     & 1   \\ \cmidrule(l){2-8}
      & Our Work && \SI{32}{\bit}              & 98.6  &  238   & 98.45 & 7  \\
      & Our Work  && \SI{8}{\bit}               & 98.51 &  70   & 41.63 & 3  \\ 
      & Our Work              && \SI{4}{\bit}               & 98.65 & 45  & 12.2  & 1  \\
      & Our Work              && \SI{2}{\bit}               & 98.46 & 29  & 8.78  & 1  \\ \midrule
                       & LoLa &\cite{BGE19}    & 8-\SI{9}{\bit} & 74.1          & 93                & \multicolumn{1}{c}{-}     & 4                \\
\multirow{5}{*}{\rotatebox{90}{CIFAR}} & HCNN&\cite{BJLMJTNAC21}    & \SI{8}{\bit}              & 77.55         & 218               & \multicolumn{1}{c}{-}     & 10               \\ \cmidrule(l){2-8}
      & Our Work             && \SI{32}{\bit}                      & 73.28 & 614 & 12801 & 30 \\
      & Our Work            & & \SI{8}{\bit}               & 73.04 & 205 & 4267  & 10 \\
      & Our Work            & & \SI{4}{\bit}               & 72.49 & 153 & 3413  & 8  \\
      & Our work            & & \SI{2}{\bit}               & 69.14 & 124 & 2560  & 6  \\ \bottomrule
\end{tabular}
\end{table}

Table~\ref{tab_QAT} compares our QAT network to earlier works. 
Notable is that the CryptoNets and HCNN implementations use PTQ techniques, but no QAT techniques. 
Using QAT, we can quantise the weights to as low as \SI{2}{\bit}, giving the network a much lower FIW with minimal to no drop in accuracy. Compared to a full precision network, i.e. quantising the parameters to \SI{32}{\bit} integers, the FIW is reduced with a factor of 8.2 for the MNIST network and a factor of 5 for the CIFAR network. Compared to the numbers presented by HCNN, our smallest networks have a 33\% and 43\% smaller final integer width for MNIST and CIFAR, respectively, while boasting similar accuracy.

\section{Evaluation}

In this section, we evaluate our newly developed quantised neural networks by implementing them using the Pyfhel~\cite{pyfhel} library, which is a software package that provides python-bindings for Microsoft's SEAL library~\cite{sealcrypto}. An encrypted inference is executed using the integer-based BFV scheme. All of our tests are run using Python 3.9.13, Brevitas 0.7.1, Pyfhel 3.3.1 (using SEAL 3.7) on an Intel Xeon Silver 4208 CPU. 

One of the most compelling optimisations in certain HE schemes was the introduction of batching or packing, as described by \citet{SIMD}. It provides a way to pack multiple plaintext messages into a single ciphertext as if it were a vector of plaintexts. In our implementation, we use batching to pack each input channel into a single ciphertext. A single ciphertext is used for a (black-and-white) MNIST image, and three ciphertexts are needed for an (RGB) CIFAR image. 


Due to batching, we cannot implement a dot-based matrix-vector multiplication since we need access to the individual elements of a ciphertext. Therefore rotation-based versions of each neural network layer are implemented based on previous works.
\citet{CHET} propose an algorithm to calculate a single convolution kernel on a subset of the input data. 
We adapted the algorithm further to enable us to apply an input kernel to a complete channel simultaneously.  
As proposed by \citet{JVC18}, a rotation-based algorithm executes a matrix-vector multiplication for the dense layer. This algorithm will perform the multiplication using only vector addition, multiplications and rotations. 

When converting to an almost binary size (\SI{2}{\bit}), extra sparsity is introduced, which we use to reduce the latency. We check for a zero vector before encoding the weight during the encrypted inference. If there is one, all the associated operations using this vector can be omitted. This results in a speedup of 28\% between a \SI{4}{\bit} and \SI{2}{\bit} network.  

\subsection{Homomorphic encryption parameter selection}



To determine and select suitable HE parameters, we first analyse the final integer width that determines whether we need multiple instances. The SEAL library limits the maximum size of the plaintext modulus to \SI{60}{\bit} for performance reasons.
Due to the outcomes of our QAT experiments, we need to represent larger plaintext spaces and use a residue numeral system (RNS). 
An overview of the used HE parameters is given in Table~\ref{he_params}.

\begin{table}[!ht]
\centering
\caption{Used HE parameters}
\label{he_params}
\begin{tabular}{@{}llccP{3.5cm}@{}}
\toprule
Network & Quantisation   & N        & $\log\;q$ & Plaintext modulus \\ \midrule
\multirow{3}{*}{\rotatebox{90}{MNIST}}   & \SI{8}{\bit}   & $2^{14}$ & 389      & 35184371138561,      \\
  & \SI{4}{\bit}   & $2^{14}$ & 389      & 35184371138561     \\
   & \SI{2}{\bit}   & $2^{14}$ & 389      & 1073643521        \\ \midrule
\multirow{6}{*}{\rotatebox{90}{CIFAR}}   & \SI{8}{\bit}   & $2^{15}$ & 825      & Same as \SI{4}{\bit} + \newline 8257537, 6946817 \\
   & \SI{4}{\bit}   & $2^{15}$ & 825      & Same as \SI{2}{\bit} + \newline 5767169, 6750209 \\
   & \SI{2}{\bit}   & $2^{15}$ & 825      & 1376257, 1769473, 2424833, 2752513, 3604481, 3735553
                  \\ \bottomrule
\end{tabular}
\end{table}


\subsection{Results}

We report the sequential times for various quantisation on the right in Table~\ref{tab_QAT}. To account for the number of instances, the `sequential time' is given, corresponding to the total time when each instance is executed sequentially and reflecting the use of computing resources. 
For the CIFAR network, the work of \citet{BJLMJTNAC21} uses ten instances, each possessing a plaintext size of around \SI{21}{\bit}. Using the same sizes, our smallest network (\SI{2}{\bit}) only requires six instances.

The MNIST architecture's smallest network is 80\% faster than the best \SI{8}{\bit} PTQ network. This is due to the smaller FIW and because it uses the additional sparsity of the weights. As for the CIFAR architecture, we obtain a 40\% speedup compared to the \SI{8}{\bit} PTQ network, which is equal to the quantisation used by HCNN.  

\section{Conclusion}

The absence of a division operation in some fully homomorphic encryption schemes implies that variables keep growing during computations. In this work, we tested two main quantisation techniques to reduce the size of the internal variables, which in turn affects computation cost. We first looked at the limitations of post-training quantisation and showed that there is a lower limit to the quantisation (in our case \SI{8}{\bit}) before the accuracy significantly drops. To further reduce the variable sizes, we developed a quantisation-aware training framework. We reduced the final integer width with 33\% for MNIST and 43\% for CIFAR, compared to the state-of-the-art HCNN architecture. In our experiments, the quantisation aware training, allowing for reducing the network weights up to \SI{2}{\bit}, boasts an 80\% and 40\% speedup for the MNIST and CIFAR network, respectively, over typical \SI{8}{\bit} weights obtained with post-training quantisation.

\newpage
\bibliographystyle{IEEEtranN}
\bibliography{ref1}

\end{document}